\documentclass[aps,superscriptaddress,twocolumn,twoside,floatfix,prl,nofootinbib,a4paper]{revtex4-1}
\usepackage{graphicx,amsmath,amssymb,color}
\usepackage[normalem]{ulem}
\usepackage{amsfonts}
\usepackage[ocgcolorlinks,colorlinks=true,linkcolor=blue,citecolor=red]{hyperref}
\usepackage{latexsym}
\usepackage{amsfonts}
\usepackage{mathrsfs}
\usepackage{natbib}
\usepackage{color,verbatim}
\DeclareMathAlphabet{\mathrsfs}{U}{rsfs}{m}{n}
\DeclareMathAlphabet{\mathpzc}{OT1}{pzc}{m}{it}
\DeclareMathAlphabet{\matheus}{U}{eus}{m}{n}
\DeclareMathAlphabet{\mathbbold}{U}{bbold}{m}{n}

\newcommand{\ba}{\begin{eqnarray}}
\newcommand{\be}{\begin{equation}}
\newcommand{\ee}{\end{equation}}

\newcommand{\ea}{\end{eqnarray}}
\newcommand{\ban}{\begin{eqnarray*}}
\newcommand{\ean}{\end{eqnarray*}}

\begin{document}

\title{ Quantum Clock Synchronization with a Single Qudit}

\author{Armin Tavakoli}
\affiliation{Department of Physics, Stockholm University, S-10691 Stockholm, Sweden}

\author{Ad\'an Cabello}
\affiliation{Departamento de F\'{\i}sica Aplicada II, Universidad de Sevilla, E-41012 Sevilla, Spain.}

\author{Marek \.Zukowski}
\affiliation{Instytut Fizyki Teoretycznej i Astrofizyki, Uniwersytet
Gda\'{n}ski, PL-80-952 Gda\'{n}sk, Poland.}

\author{Mohamed Bourennane}
\affiliation{Department of Physics, Stockholm University, S-10691 Stockholm, Sweden}


\begin{abstract}
Clock synchronization for nonfaulty processes in multiprocess networks is indispensable for a variety of technologies. A reliable system must be able to resynchronize the nonfaulty processes upon some components failing causing the distribution of incorrect or conflicting information in the network. The task of synchronizing such networks is related to detectable Byzantine agreement (DBA), which can classically be solved using recursive algorithms if and only if less than one-third of the processes are faulty. Here we introduce a nonrecursive quantum algorithm that solves the DBA and achieves clock synchronization in the presence of arbitrary many faulty processes by using only a single quantum system.

\end{abstract}
\maketitle


{\em Introduction.---}In many multiprocess networks, including data transfer networks, telecommunications networks, and the global positioning system, the individual processes need to have clocks that must be synchronized with one another \cite{SWL, LAK}. To this purpose, individual processes' clocks must periodically be resynchronized. This motivates the need for clock synchronization algorithms which work despite the faulty behavior by some of the processes. Faulty behavior can occur due to a variety of causes, including crashing, transmission failure, and distribution of incorrect or inconsistent information in the network \cite{ByzClockSync}. A clock synchronization algorithm should achieve the following tasks: C1) For any given instant, the time of all nonfaulty processes' clocks must be the same. This is necessary, but not sufficient, since simply stopping all clocks at zero satisfies C1. We therefore need to assume that a process' logical clock also keeps the rate of its corresponding physical clock. In addition, synchronizing may cause further errors, so we require that: C2) There is a small bound on the amount that a process' clock is changed during synchronization \cite{LampSync}.

Reliable clock synchronization algorithms can be complicated. To simplify the problem we shall work under the following assumptions \cite{LampSync}: A1) Initially, all clocks are synchronized to the same value. Physical clocks typically do not keep perfect time but drift to respect one another. This motivates the following assumption: A2) All nonfaulty processes' clocks run at one second in clock time per second in real time. A general problem arises from the clocks continuously changing during the synchronization procedure. Unless the synchronization algorithm is very fast, this will cause problems. This motivates our last assumption: A3) A nonfaulty process can read the time difference between the clock of another process and its own.

A method to achieve synchronization is to use interactive consistency algorithms (ICAs) in which all nonfaulty processes reach a mutual agreement about all the clocks \cite{LampSync}. A ICA should satisfy that, for every process $p$: (1) Any two nonfaulty processes obtain the same value of process $p$'s clock, even if $p$ is faulty. (2) If $p$ is nonfaulty, then every nonfaulty process obtains the value of $p$'s clock.quit

The conditions for ICAs make them suitable for the task of fault tolerant synchronization. For most applications it is sufficient to consider a scenario called detectable Byzantine agreement (DBA) or detectable broadcast \cite{FGM01, FGHHS02}.
In this case, it is required that: (i) either all nonfaulty processes obtain the same value or all abort, and (ii) if process $p$ is nonfaulty, then either every nonfaulty process obtains the same value or aborts. By ``abort'' we mean treating the value as undefined and exiting the protocol.

Classical ICAs can only achieve fault tolerant synchronization through DBA if less than one-third of the processes are faulty \cite{LampSync} and agreement is achieved by majority voting using a recursive algorithm, called $OM(n)$, where $n$ is the number of faulty processes. The $OM(n)$ algorithm works as follows. We label the processes as $P_k$, with $k=1,2,\ldots,m$.
If $n=0$, then $P_1$ distributes its value to every other process. Every process uses the value received from $P_1$ and, in case no value is obtained, uses $0$. If $n>0$, then $P_1$ distributes its value to every other process. For $k=2,\ldots,m$, let $x_k$ denote the value obtained by $P_k$ from $P_1$. If $P_k$ receives no message, then let $x_k=0$. $P_k$ acts as $P_1$ in algorithm $OM(n-1)$ by distributing $x_k$ to the remaining $m-2$ processes. For every $k$ and $\forall j\neq k$, let $x_j$ be the value received by $P_k$ from $P_j$ using $OM(n-1)$, and in case no value was received $x_j=0$. $P_k$ decides on the value obtained from the median of $(x_1,\ldots,x_m)$. Thus, $OM(n)$ requires $O(m^{n+1})$ transmitted messages to solve the task.

The DBA is an example of a communication task for which quantum resources can provide a solution, while classical tools cannot. Nevertheless, the sepcial case of DBA in a three process network where one is faulty, has been solved using quantum methods based on three-qutrit singlet states \cite{FGM01, Cabello02}, four-qubit entangled states \cite{Cabello03b,GBKCW07}, and three
\cite{FGHHS02} or two \cite{IG05} pairwise quantum key distribution (QKD) channels.

Interestingly, later works have shown that there are quantum solutions for certain communication complexity problems and secret sharing tasks which do not require entanglement, but, instead, sequential communication of a single quantum system \cite{STBKZW05, TSBBZW05}. These protocols have been shown to be much more resistant to noise and imperfections, and significantly more scalable than protocols based on entanglement.

In this paper, we introduce a quantum ICA that solves the DBA and achieves clock synchronization in the presence of an arbitrary number of faulty processes, with only one single round of message passing per process independently of the number of faulty processes, utilizing only a single quantum system.


\begin{table*}[htb]
\caption{\label{tab1} Once $P_k$  receives all messages and lists from all other processes, it will study the obtained lists and messages and compare to its own list $l_k$. Depending on the consistency between obtained and private data $P_k$ will act according to table below. Notation $\{m_{j,k},l_{j,k}\} \cong l_k$ means that $m_{j,k}$ and $l_{j,k}$
are found to be consistent with $l_k$ whereas $\not\cong$ means
``inconsistent with.'' The symbol $\bot$ means ``I have received
inconsistent data.'' By $\mathbb{M}_k$ we denote some non-empty subset of $\{1,\ldots,m\}\setminus\{k\}$.}
\begin{ruledtabular}
{\begin{tabular}{c|l|l}
& {\rm local analysis of all data received by  $P_k$} & {\rm decision of  $P_k$ on the value $V_k$} \\
\hline
(iia) & $\forall j\in \mathbb{N}_m\setminus\{k\}, \hspace{1 mm} \{m_{j,k},l_{j,k}\}\cong l_k$ and all messages are equal & {\rm $V_k =m_{1,k},$ no faulty process} \\
(iib) & $\forall j\in \mathbb{N}_m\setminus\{k\}, \hspace{1 mm} \{m_{j,k},l_{j,k}\}\cong l_k$ and {\em not }all messages are  equal & as  $P_1$ is faulty, $V_k=abort$ \\
(iic) & $\forall j\in\mathbb{M}_k, \hspace{1 mm} \{m_{j,k},l_{j,k}\}\ncong l_k$ and $\forall j\notin \mathbb{M}_k, \hspace{1 mm} \{m_{j,k},l_{j,k}\}\cong l_k$ & {\rm $V_k=m_{j,k}$, for $j\notin \mathbb{M}_k$, as the other  $P_j$'s  are faulty} \\
(iid) & $\forall j\in \mathbb{M}_k, \hspace{1 mm} \{m_{j,k},l_{j,k}\}\cong l_k$ and $\bot$ $\forall j\notin \mathbb{M}_k$ & $V_k=m_{j,k}$, although $P_1$ could be faulty\\
(iie) & $\forall j\in \mathbb{M}_k, \hspace{1 mm} \{m_{j,k},l_{j,k}\}\cong l_k$, but with   unequal messages, and $\bot$ from  $\forall j\notin \mathbb{M}_k$ & {$V_k=abort$, at least $P_1$ is faulty}
\end{tabular}}
\end{ruledtabular}
\end{table*}

In order to solve the DBA problem, the $m$ processes need to share data in the form of lists $l_k$, of numbers subject to specific correlations, and the distribution must be such that the list $l_k$ held by process $P_k$ is known only by $P_k$. Quantum mechanics provides methods to generate and securely distribute such data, here we shall seek for one which is simple, efficient, and easily extendible to an arbitrary number of processes.
We assume that all processes can communicate with one another with oral messages by pairwise authenticated error-free classical channels and pairwise authenticated quantum channels.


{\em Correlated lists and their use.---}The initial stage of the quantum protocol is to distribute lists $l_{k}$, for $k=1,\ldots,m$, each of them available only to process $P_k$. All lists have to be of the same length $L$ and are required to satisfy the property that if $N=0$ (or $1$) is at position $j$ in $l_{1}$, then $0$ (respectively, $1$) is at position $j$ in lists $l_k$ for $k=2,\ldots,m$ (i.e., they are perfectly correlated). However, if $N\in \{2,\ldots,m-1\}$ is at position $j$ in $l_{1}$, then the sum of numbers at positions $j$ in lists $l_k$ for $k=2,\ldots,m$ equals $m-N$, and all elements in these lists are either $0$ or $1$. Given an $N$, all the possible combinations of binary numbers satisfying the condition are uniformly probable.

Note that, on one hand, $P_1$ has information about at which positions the lists of all other processes the values are perfectly correlated, and at which positions they are random bits, with the property that their sum is anticorrelated with the value, $N\geq 1$, in $l_k$. On the other hand, the holder of one the lists $l_k$, with $k=2,\ldots,m$, has no information whatsoever on whether the lists are correlated at a given position or not.

Once the processes have these lists, they can use them to achieve mutual agreement and solve the DBA by applying the algorithmic part of the protocol, which we shall call $QB(n,m)$. The special case, $QB(1,3)$, reproduces the protocol in \cite{GBKCW07}.

(1) $P_1$ sends bit-valued messages to all processes. The message sent to process $P_k$ will be denoted by $m_{1,k}$. Together with each message, $P_1$ sends a list $l_{1,k}$ of all of the positions in $l_{1}$ in which the value $m_{1,k}$ appears. If $P_1$ is nonfaulty all lists and messages are identical. The full information which $P_k$ receives from $P_1$ will be denoted by $\{m_{1,k},l_{1,k}\}$.

(2) The receiving processes $P_k$ analyze (singlehandedly) the obtained lists and messages. If the analysis of $P_k$ shows that $l_{1,k}$ is of appropriate length (i.e., about $L/m$) and $\{m_{1,k},l_{1,k}\}$ is consistent with $l_k$ at all positions, then if $P_k$ is nonfaulty, it conveys $\{m_{1,k},l_{1,k}\}$ to all other processes $P_{k\neq1}$. A faulty process sends a flipped bit value of the message with a whatever list it chooses. The full information which $P_j$ receives from $P_k$ will be denoted by $\{m_{k,j},l_{k,j}\}$.

A nonfaulty $P_k$ will also decide on the final bit value it adopts $V_k$. This is $m_{1,k}$, unless messages from the other processes force it to decide that $P_1$ is faulty. However, if $\{m_{1,k},l_{1,k}\}$ is not consistent with $l_k$, then $P_k$ immediately ascertains that $P_1$ is faulty and relays to other processes neither $0$ nor $1$ but $\bot$, meaning ``I have received inconsistent data.''

(3) Once all messages have been exchanged between $P_2,\ldots,P_m$, each process considers the obtained data and acts according to the instructions in Table~
\ref{tab1}. The overall aim is, if $P_1$ is nonfaulty, to have the same value of $V_k$ for all nonfaulty processes, or all of them aborting.


{\em Quantum protocol for distributing lists $l_k $.} All processes are equipped with devices which can unitarily transform qudits. In addition, $P_1$ has a source of {\em single qudits of dimension} $m$ and the last process, $P_m$, has {\em additionally} a measurement device. The protocol runs as follows (for an illustration, see Fig.~\ref{Fig1}):

\begin{figure}
\begin{center}
\includegraphics[scale=0.36]{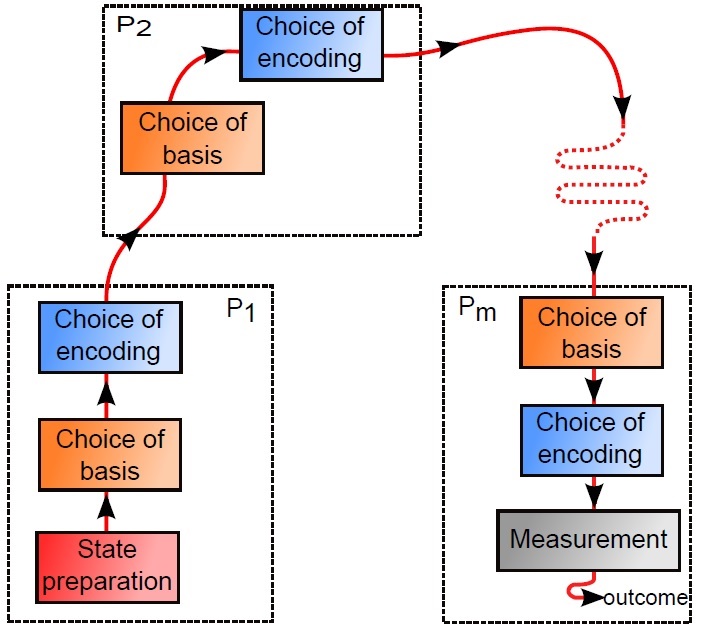}
\caption{Scheme of the quantum protocol for the distribution of the correlated lists. $P_1$ prepares a uniform $d$-level superposition state, makes a choice of basis and encoding, and forwards the qudit to $P_2$ which applies a choice a basis and encoding and forwards the qudit to $P_3$. Processes $P_3,...,P_m$ act in analogy with $P_2$. Finally $P_m$ projects the state onto the initial state prepared by $P_1$ and if the outcome is $1$ the round is treated as valid.}
\label{Fig1}
\end{center}
\end{figure}

(I) $P_1$ prepares the state
\begin{equation}
|\psi_0\rangle=\frac{1}{\sqrt{m}}\sum_{j=0}^{m-1}|j\rangle.
\label{state}
\end{equation}

(II) $P_1$ randomly chooses the ``encoding basis'' from $m$ different options $U_{0},...,U_{m-1}$ and labels the choice $c_1$. Having chosen the $c_1$'st encoding basis, process $P_1$ applies the following unitary transformation to the qudit:
\begin{equation}
U_{c_1}=|0\rangle\langle 0|+\sum_{k=1}^{m-1} \omega^{c_1}|k\rangle\langle k|,
\end{equation}
where $\omega=e^{i\frac{2\pi }{m}}$. From the interferometric point of view, applying $U_{c_1}$ introduces a phase-shift of $-2\pi c_1/m$ in the first beam.

(III) After that, $P_1$ randomly chooses a value $N_1$ in the set $\{0,1,\ldots,m-1\}$ and encodes $N_1$, by applying the following unitary transformation:
\begin{equation}
 U(N_1)=\sum_{j=0}^{m-1}\omega^{jN_1}|j\rangle\langle j|.
\end{equation}
Afterwards, the qudit is sent to $P_2$.

(IV) $P_2$, in the same manner as $P_1$, choses a $c_2\in\{0,...,m-1\}$ and applies a the unitary $U_{c_2}$ corresponding to choice of encoding basis.

(V) Next, $P_2$ randomly chooses a value $N_2$ in the set $\{0,1\}$. If $N_2=0$, no action is taken, i.e., $P_2$ applies the transformation $U(N_2=0)=\mathbf{1}$. If $N_2=1$, then $P_2$ applies $U(N_2=1)$ and then sends the qudit to $P_3$.

(VI) $P_3,\ldots,P_m$ consecutively repeat the same procedure as $P_2$ with independent choices of basis and encoding their respective random values $N_3,\ldots, N_m$.

(VII) In addition, $P_m$ measures the qudit using a device which distinguishes the state $|\psi_0\rangle$ from any set states orthogonal to it.

(VIII) If $P_m$ obtains $|\psi_0\rangle$, then the processes consecutively reveal their encoding bases (but not their values $N_k$) in reverse order: First $P_m$ and last $P_1$. If it turns out that the sum of the basis choices modulo $m$ equals zero, then the run is
treated as a valid distribution of the numbers $N_k$ at the same position in the private lists $l_k$.

The protocol distributes the numbers in the required way because all the unitary operators are diagonal and, therefore, commute. Additionally, if $\sum_{k=1}^{m}c_k=0\mod{m}$ then
\begin{equation}
\prod_{k=1}^{m}U_{c_k}=\mathbf{1},
\end{equation}
and, if $\sum_{k=1}^{m} N_k=0 $, modulo $m$, then
\begin{equation}
 \prod_{k=1}^{m}U(N_k)=\mathbf{1}.
\end{equation}
Whenever this condition is not satisfied, the final state of the system is orthogonal to $|\psi_0\rangle$ and will therefore never be an outcome of $P_m$'s measurement.


{\em Clock synchronization.---}Fault tolerant clock synchronization is one possible adaption of our method to achieve DBA. However, in this case, a problem arises from clocks ticking during the synchronization procedure. This is solved by exploiting assumption A3: Instead of sending a number, the processes send their clock differences to each other. In the classical case, we achieve clock synchronization by running the algorithm $OM(1)$ $m$ times, sending clock differences instead of the binary values, and analogously for $OM(n)$ \cite{LampSync}. In analogy with the classical case, the processes send clock differences also in the quantum case, exploiting the fact that the clock differences can be decomposed into binary strings up to arbitrary accuracy agreed upon in advance. We run $QB(n,m)$ $m$ times in such a way that for each run a new processes takes the roll of $P_1$ in $QB(n,m)$. More explicitly, $P_y$ reads the clock difference $\Delta_{xy}$ between its own clock and the clock of $P_x$. If $P_y$ is nonfaulty it will relay $\Delta_{xy}$ to $P_z$ but if $P_y$ is a faulty process, it can arbitrarily change $\Delta_{xy}$ before sending it. If $P_y$ relays the value obtained from $P_x$ to $P_z$, then $P_z$ knows the time difference between $P_x$ and $P_y$. Also, since $QB(n,m)$ is ran $m$ times, $P_z$ will also obtain $\Delta_{yz}$ from $P_y$ and thus $P_z$ knows that $P_y$ is claiming that the time difference between $P_x$ and $P_z$ is $\Delta_{xy}+\Delta_{yz}$, which can then be compared to $\Delta_{xz}$ obtained directly from $P_x$.


{\em Comparison with the other solutions.---}The correlated lists needed for achieving DBA can be distributed by other means than with the single-qudit protocol. Successful distribution can be achieved by the process $P_m$ sharing a QKD channel with every other process. $P_m$ uses a QKD protocol, e.g., BB84 \cite{BB84} to distribute numbers such that (1) $P_m$ and $P_1$ share a string $K_{1,m}=k_{1,m}^1 \ldots k_{1,m}^L$, where $k_{1,m}^j \in \{0,\ldots,m-1\}$. (2) For every $l=2,\ldots,m-1$, $P_m$ and $P_l$ share a string $K_{l,m}=k_{l,m}^1 \cdots k_{l,m}^L$ such that $k_{l,m}^j\in \{0,1\}$. (3) For a given $j$, the lists satisfy $(\sum_{l=1}^{m} k_{l,m}^j)_{\rm{mod}\,m}=0$. (4) None of $P_2,\ldots,P_{m-1}$ have any information about a particular list element of any other process. (5) Whenever $P_1$ receives an element $k_{1,m}^j\geq 2$, $P_1$ has no information on the bit value of $k_{l,m}^j$ for $l=2,\ldots,m$, and whenever $P_1$ receives $k_{1,m}^j=p\in\{0,1\}$, $P_1$ knows that $k_{l,m}^j=p$ for all $l=2,\ldots,m$. All QKD channels except that shared between $P_1$ and $P_m$ transmit bit values. In order to transmit elements of $\{0,\ldots,m-1\}$ to $P_1$, the numbers must be encoded into $\lceil \log_2\left(m\right)\rceil$ qubits. One additional requirement that has to be made for solving the DBA using the QKD distributed lists is that $P_m$ is not required to convey any lists. This is necessary since $P_m$ has full knowledge about the lists of all other processes and therefore easily could cheat. Instead, $P_m$ may announce the message it received from $P_1$, and if any inconsistency is noted by $P_2,\ldots,P_{m-1}$, then $P_m$ will change its final value if the other processes convince $P_m$ of them being nonfaulty.

There is also a number of proposed solutions to the DBA considering three processes where one is faulty. The first one, proposed in Ref.~\cite{FGM01}, relies on the three qutrit entangled Aharonov state. The goal is to distribute lists given by all permutations of the elements of the set $\{0,1,2\}$, i.e., (0--1--2, 0--2--1, 1--0--2, 1--2--0, 2--0--1, and 2--1--0). Generalization to $m$ parties along the lines of \cite{FGM01} would require the usage of multipartite $m$-level entanglement, provided by the state
\begin{equation}
\label{state}
|\kappa_m\rangle=\frac{1}{\sqrt{m!}}\sum_{\overline{i}= \sigma(S_m)} (-1)^{N(\sigma(S_m))}|i_1,\ldots,i_m\rangle,
\end{equation}
where $\overline{i}=\{i_1,\ldots,i_n\}$, $S_m=\{0,\ldots,m-1\}$ and $N(\sigma(S_m))$ is the parity of the permutation of $S_m$. Already for the simplest case of $m=3$, this approach requires the preparation of a very complex state which, to our knowledge, has not yet experimentally realized. However, for the three process case, it has been pointed out in \cite{IG05} that the distribution of the lists can be realized without the state (\ref{state}), by utilizing two separated QKD channels. With small modification for the $m$ process setting, distribution of the lists is achieved with $m-1$ QKD channels. However, to encode the entire space provided by $S_m$, the QKD requires $\lceil\log_2\left(m\right)\rceil$ qubits. If the efficiency of a detector $\eta$ is not perfect and the QKD is performed with single qubits using von Neuman measurements, successful distribution occurs only with probability $\eta^{(m-1)\lceil\log_2\left(m\right)\rceil}$. Typically, the classical part of the protocol in \cite{FGM01} and its possible generalizations scale rapidly with the number of processes. It is required that $m!$ different types of lists are distributed. However, a solution to the three party DBA exploiting four-qubit entanglement provides a simpler classical part of the protocol: the number of different lists is lowered from six to four \cite{GBKCW07}.

The general $m$ process protocol presented in this paper generalizes the protocol in \cite{GBKCW07} and requires $2^{m-1}$ different types of lists. As emphasized earlier, the distribution of the required lists can be achieved both with single-qudit and with $m-1$ QKD channels. Using QKD channels, only one channel needs to transmit all elements in $S_m$ while the remaining $m-2$ channels only transmit bit values. In the presence of nonperfect detectors, successful distribution occurs with probability $\eta^{m-2+\lceil\log_2\left(m\right)\rceil}$. However, in the single-qudit approach only one single detection is needed and, therefore, successful distribution of the lists occur with probability $\eta$ independently of $m$. The single-qudit protocol is highly scalable, both in terms of success probability with inefficient detectors and requirements on the classical lists.


{\em Conclusions.---}We have presented a single-qudit protocol which provides an efficient solution to an important multiparty communication problem: It solves DBA and achieves clock synchronization in the presence of arbitrary many faulty clocks. In principle, our quantum algorithm is not limited to the case of clock synchronization, it can with small modification be used for other tasks requiring oral message interactive consistency. Interestingly, our algorithm works by transmitting a single qudit among the parties rather than by distributing a quantum entangled state among them. This makes the protocol much more practical, as single qudits can be experimentally realized easily in many ways. For example, using unbiased multiport beamsplitters \cite{ZZH97} or time-bin \cite{MRTSZG02}. Compared to schemes based on several QKD channels, the single-qubit protocol is more scalable and robust against detection inefficiencies. This results shows that single-qudit quantum information protocols are interesting beyond QKD \cite{CBKG, CANS, Svozil09b} and random number generation \cite{Svozil09a, UZZWYDDK13}, and should stimulate experimental implementations and further research in quantum information protocols.

{\bf Acknowledgements:} This project was supported by the Swedish Research Council, ADOPT, the Project No.\ FIS2011-29400 (MINECO, Spain) with FEDER funds,
the FQXi large grant project ``The Nature of Information in Sequential Quantum Measurements,'' MNiSW Grant
No. IdP2011 000361 (Ideas Plus) and Foundation for Polish Science TEAM project co-financed by
the EU European Regional Development Fund. The work is an extension of the first sketch presented in \cite{QUANT-PH}.\\



\newpage

\end{document}